\documentclass[12pt]{article}
\usepackage{epsfig}
\usepackage{cite}
\usepackage{./Macros/mcite}
\usepackage{url}

\usepackage{array,tabularx,epsfig,mathrsfs,graphicx,rotating}
\usepackage{ifthen}
\usepackage{fancybox}
\usepackage{amsfonts}

\newcommand{\beq}{\begin{equation}}
\newcommand{\eeq}{\end{equation}}

\newcommand{\Fc}{\mathcal{F}}
\newcommand{\Za}{\mathrm{Z_1}}
\newcommand{\Zb}{\mathrm{Z_2}}
\newcommand{\Zn}{\mathrm{Z_n}}
\newcommand{\F}{\mathrm{F}}

\chardef\til=126
\newcommand{\mev}{{\,\mathrm{MeV}}}
\newcommand{\gev}{{\,\mathrm{GeV}}}

\begin{document}

\clearpage
\pagestyle{empty}
\setcounter{footnote}{0}\setcounter{page}{0}%
\thispagestyle{empty}\pagestyle{plain}\pagenumbering{arabic}%

\hfill  DESY 06-138 

\hfill  ANL-HEP-PR-06-61 
 
% \hfill Version 2.4

\vspace{2.0cm}

\begin{center}

%%%%%%%%%%%%%%%%%%%%%%%%%%%%%%%%%%%%%%%%%%%%%%%%%%%%%%%%%%%%%%%
{\Large\bf
Regularities in hadron systematics, Regge trajectories
and a string quark model
\\[-1cm] }
%%%%%%%%%%%%%%%%%%%%%%%%%%%%%%%%%%%%%%%%%%%%%%%%%%%%%%%%%%%%%%%

\vspace{2.5cm}

{\large S.V.~Chekanov $^a$
\footnote[1]{
On leave from
HEP Division, Argonne National Laboratory,
9700 S.Cass Avenue,
Argonne, IL 60439
USA
}
and B.B.~Levchenko $^b$

\begin{itemize}
\itemsep=-1mm
% \leftskip-0.5cm

\normalsize
\item[$^a$]
\small
DESY Laboratory, 
Notkestrasse 85, 22607, Hamburg, 
Germany

\normalsize
\item[$^b$]

\small
Skobeltsyn Institute of Nuclear Physics, Moscow State University, \\ 
119992 Moscow, Russian Federation

\end{itemize}
}

\normalsize
\vspace{1.0cm}

% Short title: Baryon spectra, Regge trajectory, hadron systematics, diquark model 

% put line numbers
% \linenumbers

\vspace{0.5cm}
\begin{abstract}
An empirical principle for 
the construction of a linear relationship between 
the total angular momentum and squared-mass
of baryons is proposed. 
In order to examine linearity of the trajectories, 
a rigorous least-squares regression analysis was performed. 
Unlike the standard Regge-Chew-Frautschi approach,
the constructed trajectories do not have non-linear behaviour.
A  similar regularity may exist for lowest-mass mesons. 
The linear baryonic trajectories are well described by a semi-classical picture
based on a spinning relativistic string with tension.
The obtained numerical solution of this model
was used to extract the (di)quark masses.
\end{abstract}

\end{center}

\newpage
\setcounter{page}{1}

% put line numbers
% \linenumbers

%%%%%%%%%%%%%%%%%%%%%%%%%%%%%%%%%%%%%%%%%%%%%%%%%%%%%%%%%%%%%%%%%%
\section{Introduction}
%%%%%%%%%%%%%%%%%%%%%%%%%%%%%%%%%%%%%%%%%%%%%%%%%%%%%%%%%%%%%%%%%%

It has been accepted for a long time that hadrons from the same family 
lie on Regge trajectories 
(the so-called Chew-Frautschi conjecture \cite{Chew:1961ev, *Chew:1962eu, *chew}), i.e.
\begin{equation}
J  = \alpha(0) + \alpha^{'} M^2,
\label{eq1}
\end{equation} 
where $J$ is the total angular momentum and $M$ is the mass of a hadron. 
The intercept $\alpha(0)$ depends on hadron type,
but the slope  $\alpha^{'}$ is approximately the same for all hadrons.
Such a relationship between $J$ and $M^2$, also
known as the principle of exchange degeneracy, 
is usually interpreted  as a
manifestation of the linear potential of the strong forces between 
constituent quarks. Recently this picture was discussed 
in terms of a relativistic diquark model based on a spinning string with 
a constant tension \cite{sim1,*sim2,solo,selem} (see also references in \cite{Inopin:1999nf}).

Experimentally, the validation of Eq.~(\ref{eq1}) 
remains to be a difficult problem since the experimental data are scarce. 
According to a  recent classification of hadrons on the   
($J,M^2$) plane\cite{anisovich}, the overwhelming majority 
of the trajectories are supported by  a few data points only.
For mesons, there exist five trajectories with three data points,
while other trajectories were hypothesized from two or one data
points. 
For baryons, only one trajectory was constructed from four data points and 
five trajectories were supported by three data points. 
Other trajectories were hypothesized 
from the study of either one or two data points. 
Recently, it has been noted that only $9\%$ ($14\%$) of all trajectories 
in the mesonic (baryonic) sector are linear~\cite{inopin}. 
A non-linear character of the Regge trajectories has also been 
pointed out in \cite{tang,desgrolard,selem}.

In this article we are not going to scrutinize  
the linearity of the Regge trajectories; 
it is rather clear that the vast majority of such 
trajectories are indeed non-linear and there is no necessity to re-analyse this
fact again.
Instead, using the most recent PDG data ~\cite{pdg},
we would like to note that it is possible to find a prescription which could allow the construction 
of the trajectories
which can be classified as
being perfectly linear and  span over a significant  number
of known  baryons. 
In order to illustrate this,
we use a rigorous weighted least-squares regression,
which is often missing in theoretical papers on this subject.
In Sect.~\ref{sec1}, we will discuss our 
principles for the  construction of  
a linear relationship between the 
hadronic  mass squared $M^2$ and the total angular momentum $J$. 
In Sect.~\ref{sec2},
our empirical observation is discussed in more details. In particular, 
we will explain what  
would  happen if the requirements proposed in
Sect.~\ref{sec1} are removed or weakened. 
In Sect.~\ref{sec3} and  appendices,
we will attempt to use our approach for validation
of a relativistic diquark model based on a spinning string,  
which is often discussed in the literature (see \cite{selem} for
a recent discussion).  Finally, we will briefly 
discuss the mesonic sector in Sect.~\ref{sec4}.

%%%%%%%%%%%%%%%%%%%%%%%%%%%%%%%%%%%%%%%%%%%%%%%%%%%%%%%%%%%%%%%%%%
\section{Linear trajectories for baryons}
\label{sec1}
%%%%%%%%%%%%%%%%%%%%%%%%%%%%%%%%%%%%%%%%%%%%%%%%%%%%%%%%%%%%%%%%%%

As a starting point in the construction of baryonic trajectories, we will consider: 
(a) only stable or strongly decaying particles from a certain family;
(b) only hadrons of the same family 
with a smallest mass, $M_{min,J}$, for a given fixed $J$ and space parity $P$;  
(c) in addition to (a) and (b), we will use only such baryons 
if there are no other baryons with smaller masses and with the opposite space parity. 
At this moment, we will use  the assumptions above without any theoretical
justification. Below we will show that, without any exception, 
such requirements are sufficient for the construction  
of perfectly linear baryonic trajectories.

As a leading principle for the selection of hadrons, 
we will ignore particles with poor evidence of existence  
i.e. with  one star in accordance with the PDG classification~\cite{pdg}. Furthermore, 
in several cases when PDG quotes only mass ranges,
we will use the central values for the masses and their 
experimental errors as they are given by 
the most accurate and recent measurements\footnote{This can be found in the 
computer files located at the PDG web 
page \cite{pdgcom}.}. 

We will attempt  to describe the mass dependence on $J$ using a linear parameterisation 
similar to Eq.~(\ref{eq1}): 
\begin{equation}
M^2_{min,J} = p_0 + p_1 J, 
\label{eq2}
\end{equation}
where $p_0$ and $p_1$ are constants.  
Hereafter we will use the  notation $M$  for $M_{min,J}$,  
unless otherwise stated.
For convenience of a linear-regression
analysis, we prefer to express $M^2$ as a function of $J$
since the angular momentum does not have experimental uncertainties. 
The linear least-squares regression analysis is described in Appendix~A.

\vspace{0.5cm}
{\bf  $N$ baryons}. First, let us consider $N$ baryons.
Figure~\ref{fig2n+} shows all PDG $N$-baryons 
on the $(J,M^2$) plane (the filled and open symbols). 
There is a clear linear trend for lowest-mass baryons with $P=+1$
shown by the filled circles (the PDG names of such baryons are indicated).
Since such baryons fully  satisfy the criteria
proposed at the beginning of this section, we will use them   
for a linear least-squares regression.
The result of this regression is shown with the solid line.
The trajectory is remarkably straight: the small value of $\chi^2/ndf$ shown 
in Fig.~\ref{fig2n+} fully 
supports the linear fit. 

To check a possible non-linearity of the $N$-baryon trajectory, a fit was performed
using a second-order polynomial function.
The inclusion of additional term to the fitting function 
leads to a small value  ($\simeq 0.005$)  of the parameter responsible for the quadratic term. 
However, the quality of such fit characterized by $\chi^2/ndf=0.15/1$ 
does not significantly improve.  
Two dashed lines shown in Fig.~\ref{fig2n+} 
indicate a $95\%$ confidence-level  region for the linear regression, assuming that
the experimental uncertainties on the masses are normally distributed. 
This illustrates the reliability of the linear relationship between the 
$M^2$  and $J$ values.
From the two tests above, it can be concluded that there is no evidence for 
non-linearity of the $N$-baryon trajectory. 

It should be noted that we do not include the $N(1990)(F17)**$ baryon at $J=7/2$ to the fit.
In principle, the quality of the fit will not suffer if the quoted PDG mass ($\simeq 1990\mev$)
is used. However, if one uses the most recent measurement given by the PDG \cite{pdgcom},
then the $N(1990)(F17)$ point will move up to the location indicated by the small symbol at 
$2086\pm 28\mev$ (Fig.~\ref{fig2n+}). 
In this case, this baryon will overlap on the $(J,M^2$) plane with   
a better established  $P=-1$ baryon $N(2190)(G17)$, 
which has a mass of $2190^{+10}_{-90}\mev$ (again in accordance
with the latest measurement). In this case, $N(1990)(F17)$ should not be
used for the fit due to the ``lowest-mass'' exclusion principle. Since there is
no any objective criteria for inclusion (or exclusion) of non-well established 
$N(1990)(F17)$ baryon to (from) the fit,
it was decided to exclude it.

The situation with the lowest-mass $P=-1$ $N$ baryons, 
shown in Fig.~\ref{fig2n+} (open symbols), 
is more complicated and will be discussed in Section~\ref{sec2}.
 
\vspace{0.5cm}
{\bf  $\Delta$  baryons}. Now let us consider the $\Delta$ baryons.
Figure~\ref{fig2delta+} shows the $(J,M^2)$ plane for all $\Delta$ baryons (filled symbols).
The filled circles show the $P=+1$ baryons with smallest 
masses at a given $J$ (their
PDG names are indicated).
It should be noted that the PDG name for the $\Delta(2000)$ 
baryon with $J=5/2$ 
is likely to be inappropriate
since  most recent experimental studies have indicated 
that its mass is either 
$1724\pm 61\mev$ \cite{del1} or $1752\pm 32\mev$ \cite{del2}.
For the linear fit, we use the former mass, which is also quoted in \cite{pdgcom}.

The linear least-squares regression for the lowest-mass $\Delta$ baryons 
with $P=+1$ is shown in Fig.~\ref{fig2delta+} by the solid line. 
The $\chi^2/ndf=6.1/4$ supports the linear fit, despite
very small experimental uncertainties on the measured masses and 
the significant range in $J$.
The quality of the linear regression 
is impressive:
The dashed lines indicate a $95\%$ confidence-level interval for the linear
regression, which is only $170\mev$ wide even for $J=21/2$.
Both multiple r-squared and adjusted r-squared are $0.9999$ and
the p-value is $6.2\cdot 10^{-9}$.
Furthermore, a possible non-linear trend was checked 
by using a second-order polynomial function. 
Such a fit has $\chi^2/ndf=5.4/3$, while
the parameter for the quadratic term was 
consistent with zero ($0.014 \pm 0.017$). 
Thus, one concludes again that there is no evidence for a non-linear behaviour.

The linear regression has only one significant residual at $J=5/2$ where the measured 
mass squared is $100\mev^2$ above the upper $95\%$ confidence-level line. 
However, as was noted above, 
the existence of the $\Delta(2000)$ baryon is not well 
settled (two PDG stars),
and its mass needs  to be determined more
accurately.
It should also be noted that only three baryons shown in Fig.~\ref{fig2delta+}, 
$\Delta(1232)$, 
$\Delta(1950)$ and $\Delta(2420)$,
are fairly well studied.
 
Using the principles outline above, it is impossible to find a sufficient 
number of
$\Delta$ baryons with $P=-1$ for the linear regression fit. 
This will be discussed in Section~\ref{sec2}. 

\vspace{0.5cm}
{\bf  $\Lambda$  baryons}.
Figure~\ref{fig2lambda} shows the $(J,M^2)$ plane for all PDG $\Lambda$ baryons 
(filled symbols).
The filled circles show the baryons with the smallest masses and with $P=-1$.
Again, such baryons are well described by the linear regression fit ($\chi^2/ndf=0.2/1$).  

Figure~\ref{fig2lambda} also shows the  $\Lambda$ baryons with smallest masses  
(their names are not indicated) and $P=+1$.
The latter baryons ($\Lambda(1115)(P01)$, $\Lambda(1890)(P03)$,  
$\Lambda(1820)(F05)$ and $\Lambda(2350)(H09)$)
cannot be used for the fit since: 1) $\Lambda(1115)(P01)$ decays weakly; 2) the 
position with $J=3/2$ is already filled with
the $\Lambda(1520)(D03)$ baryon which  has the opposite parity. We examine this
further in Sect.~\ref{sec2}.  

\vspace{0.5cm}
{\bf  Other baryons}. For $\Omega$  and $\Xi$ baryons, 
the existing experimental data are insufficient 
for the construction of the trajectories with more than two data points.
The $\Sigma$ baryons will be discussed below.

\section{Discussion}
\label{sec2}

It is rather convincing now that the method proposed 
at the beginning of Sect.~\ref{sec1} indeed works rather well:  
it allows the construction of three  perfectly
linear baryonic trajectories with more than two  data points.
The fit parameters are summarised in Table~\ref{table_fit}.  

Now, let us discuss what would happen if:  
1) weakly decaying baryons will be included in the fit as well; 
2) one considers  baryonic trajectories for a certain parity even
when there are baryons with  the opposite parity  with lower masses. 
In this context, we will also discuss hadrons with different space parity
which have been omitted from the consideration in Section~\ref{sec1}.  
We will not analyse the minimum-mass
requirement itself since,  without it, the vast majority of baryonic
trajectories will be non-linear and identical to those studied 
elsewhere~\cite{Inopin:1999nf,inopin,tang,desgrolard}.

First of all, let us include the lowest-mass weakly decaying 
$\Lambda$ state shown by the open symbol at $J=1/2$ 
in Figure~\ref{fig2lambda}. 
We will exclude the $\Lambda(1820)(F05)$ state at $J=3/2$, 
which is also indicated with the open 
symbol, since this hadron does not satisfy to the minimum-mass criteria.    
The linear fit of the remaining three baryons, 
$\Lambda(1115)(P01)$, $\Lambda(1890)(P03)$ and $\Lambda(2350)(H09)$, 
cannot be considered as a perfect one since $\chi^2/ndf = 3.2/1$.
Thus, the weakly decaying $\Lambda(1115)(P01)$ state violates the linearity.  

There is a similar situation with the $\Sigma$ baryons. Figure~\ref{fig2sigma+} shows 
the $\Sigma$ baryons with $P=+1$. There are only two points which can be used to construct
the trajectory. If the lowest-mass 
weakly decaying $\Sigma (P11)$ at $J=1/2$ is included, the data points cannot be
described by the linear fit.

The open circles in Fig.~\ref{fig2sigma+} show two $\Sigma$ baryons whose angular momenta
are unknown, but their  masses  are reasonably  well determined.
The values of the angular  momenta were hypothesized assuming that
such baryons should by close to the line determined by $\Sigma(1385)(P13)$ and $\Sigma(2030)(F17)$ 
baryons. At the same time, these new baryons 
should be above the $\Delta$-trajectory which bounds the entire baryonic spectra
at low masses.

Now let us assume that the minimum-mass requirement for a given 
parity is not essential in cases when
there exist baryons with the opposite parity but with smaller masses for a fixed $J$. 
First, let us consider the 
$N$ baryons with $P=-1$ shown in Fig.~\ref{fig2n+}(open symbols). 
For $J=3/2$ and $11/2$, one can use $N(1520)(D13)$ and $N(2600)(I111)$ 
baryons without any ambiguity.
For $J=5/2$ and $J=9/2$,
$N(1675)(D15)$ and  $N(2250)(G19)$ baryons with $P=-1$ 
overlap in masses with the $N$ baryons of 
the opposite parity, i.e. $N(1680)(F15)$ and $N(2220)(H19)$.
If one ignores the minimum-mass requirement,  then all baryons indicated by the open symbols should 
be considered for the regression. Obviously, the linear fit will fail in this case.  
However, a linear trajectory for the  $P=-1$ baryons may still exist if the $N(1675)(D15)$ and  $N(2250)(G19)$ 
states are removed from the consideration on the basis of the minimum-mass requirement. 

Now let us consider 
strongly decaying $\Lambda$ with $P=+1$ shown in Figure~\ref{fig2lambda}. 
It is evident that the data points shown  with the open symbols  
cannot be described by a linear function.
Thus, it is essential to exclude $\Lambda(1890)(P03)$ at $J=3/2$ 
from the consideration. This can indeed be done taking into account the
minimum mass requirement and noting the presence of the low-mass $\Lambda(1520)(D03)$ state. 

There is another example: $\Delta$ baryons with $P=-1$ (see Fig.~\ref{fig2delta-}). 
Such baryons cannot lie on the same line since the linear fit is characterised by 
$\chi^2/ndf = 11/3$ and a wide $95\%$ confidence interval.
This can be explained as before:  positions with $J=3/2$,  $5/2$ and $9/2$ have already
been filled with the $P=+1$ baryons with lower masses (shown in Fig.~\ref{fig2delta-} 
with small filled symbols).
Therefore, there are only two baryons left with $J=1/2$ and $11/2$, 
which is insufficient for the linear regression analysis. 

%%%%%%%%%%%%%%%%%%%%%%%%%%%%%%%%%%%%%%%%%%%%%%%%%%%%%%%%%%%%%%%%%%%%%%%
\section{Towards extraction of the diquark mass}
\label{sec3}

If our hypothesis is correct, then the present experimental
data can be used for the construction only three linear trajectories 
with more than two data points. All other baryons 
lying above such trajectories on the $(J,M^2)$ plane have 
additional contributions to their masses from 
non-pure orbital rotations. Perhaps one can use the term 
``intrinsic noise'' \cite{selem} for such states: 
all such hadrons may have 
a non-linear relationship between 
$M^2$ and $J$~\cite{Inopin:1999nf,inopin,tang,desgrolard}.

The linear trajectories proposed above can be used for the validation 
of a relativistic model in which
a hadron can be treated as a rotating flux tube (or a string) 
with a quark and a diquark at the ends.
Such a string can be characterised  by a constant
tension $T=\sigma/2 \pi$. 
For small (di)quark masses, 
an approximate solution of this model is given by the Selem-Wilczek  (SW) expression \cite{selem}:

\begin{equation}
M \simeq \sqrt{\sigma J} + \frac{2}{3} \frac{\pi^{1/2} \mu^{3/2}} {(\sigma J)^{1/4} }, 
\quad 
\mu^{3/2} = m_1^{3/2} + m_2^{3/2}, 
\label{sw} 
\end{equation}
where $m_1$ and $m_2$  are the masses of 
diquark and quark connected by a relativistic string.
The fitting function directly follows from Eq.~(\ref{sw}):
\begin{equation}
M^2 =  \sigma J + \frac{4}{3} \sqrt{\pi \mu^3} \cdot (\sigma J)^{1/4} + \frac{4}{9} \frac{\pi
\mu^3} {\sqrt{\sigma J}}.      
\label{sw1}
\end{equation}
This equation resembles the Chew-Frautschi relationship between $M^2$ and $J$ 
in the limit of small masses.

In order to test the model above, we will use the $N$ baryons shown in Figure~\ref{fig2n+}.
The fit result using the SW function Eq.~(\ref{sw1})  
is shown in Fig.~\ref{a_selem_n+} (solid line), while  
the two dashed lines
illustrate the fit sensitivity to $\mu$  
(in this case we keep the slope  parameter $\sigma$ to be the same as
for the nominal fit using  Eq.~(\ref{sw1})).
The dotted line shows the linear fit as in Figure~\ref{fig2n+}.  
The fit with the function given in Eq.~(\ref{sw1}) is strikingly good.    
The fit parameters are fully constrained by the proton mass at $J=1/2$, since this
is exactly the region with a highest sensitivity to $\mu$.  
The parameter $\sigma$ is $0.908\gev^2$, which is close to the
slope value obtained from the linear fit shown in Figure~\ref{fig2n+}.
The extracted value of the mass parameter is $\mu=0.323\gev$.              
% This model  parameter, reflecting the mass of the quark and diquark at the
%  ends of the relativistic spinning string, 
% is close to the constituent quark mass. 
% Under the assumption~\cite{selem} that the diquark mass is 
% comparable to that of a single quark, i.e. $m=m_1\simeq m_2$, one 
% obtains $m=\mu/2^{2/3}=0.203\gev$. This model parameter is close  
% to the mass of the dressed quark.
 
Similar fits have been performed using the $\Delta$ and $\Lambda$ baryons  
shown in Figures~\ref{fig2delta+} and \ref{fig2lambda}.   
For such baryons,  the fit with  Eq.~(\ref{sw1})  
did not converge:
the parameter $\mu$ had a fitting error larger than its value and the fit
had a significant sensitivity to initial fit  values.
The reason for this is rather simple: As seen from Fig.~\ref{a_selem_n+},
the last non-linear term in Eq.~(\ref{sw1}) can only be constrained by the region  $J<1$.
However, $\Delta$ and $\Lambda$ trajectories do not
have a hadron at such a small $J$. Even although such baryons cannot
constrain the mass parameter $\mu$, their linear trajectories are still consistent
with a fairly linear behaviour  of Eq.~(\ref{sw1}) at large $J$.

It should be noted that the solution given in Eq.~(\ref{sw}) was obtained in the limit
of negligible quark masses ($\mu\to 0$)  and for large $J$.
However,
the obtained values of $\mu$ and the range of $J$ used
to fit the data
may lead to a worry that the above assumptions
are not  appropriate and the terms beyond ${\cal{O}}  (\mu^{3/2})$ are necessary to consider in Eq.~(\ref{sw}).
Therefore, we  have made an attempt to solve the equations of 
the diquark string model analytically by taking 
into account high-order terms neglected
in the solution \cite{selem}.
We have  obtained a  relationship between the mass and $J$ using
a full set of ${\cal{O}}  (\mu^{3/2})$ terms and,
in addition, some terms of order ${\cal{O}}  (\mu^{5/2})$
(see Appendix~B).
It should be stressed that   
a complete ${\cal{O}}  (\mu^{5/2})$  calculation  requires 
a solution  of the appropriate quintic equation.

Figure~\ref{a_boris_n+} shows the fit using the  ${\cal{O}}  (\mu^{5/2})$ solution 
given in Eq.~(\ref{ap_10}) of Appendix~B.
The fit, shown with the dashed line,  
was performed using three free parameters, $m_1$, $m_2$ and $\sigma$. 
It was assumed that the mass of the diquark ($m_1$) is larger than that of the quark 
($m_2$) during the fit.  
The quality of the fit is fair.  It should be noted that if only two 
parameters are used for the fit, i.e. $\mu$ and $\sigma$ as
in the SW case, $\chi^2/ndf$ is smaller ($=2.2/2$). 
In is interesting to observe that $M^2$ increases when $J$ decreases.
In fact, this is an artifact of truncation of the series in Eq.~(\ref{ap_11});   
high-order terms
proportional to ${\cal{O}}  (\mu^{7/2})$ and ${\cal{O}}  (\mu^{9/2})$ have 
negatived contributions  and thus turn to reduce the increase of $M^2$ at small $J$.

It is rather clear that in order to obtain a reliable model prediction when the masses
are not too small, it is essential to find a numerical solution of Eq.~(\ref{ap_5}).    
The result of our numerical 
calculation\footnote{We used the function DZERO from the
CERNLIB FORTRAN library~\cite{cernlib} to find
a zero of a real-valued function for solving Eq.~(\ref{ap_5}). 
Then such a numerical solution was used in the fitting function using  
the MINUIT program, see Appendix~A.} 
used in a  $\chi^2$-minimisation procedure for the $N$-baryonic trajectory is 
shown in Figure~\ref{a_boris_n+} (the solid line). 
The fit was found to be excellent.

Table~\ref{table} summarizes the fit parameters obtained using different approximations.  
The numerical solution leads to very similar masses as for the
analytical ${\cal{O}}  (\mu^{5/2})$ calculation. 
Moreover, the parameter $\mu\simeq 0.314\gev$ calculated from the
extracted $m_1$ and $m_2$ is very similar to that from 
the SW fit shown in 
Figure~\ref{a_selem_n+}.    
As before, the fit can only
be constrained by the $J<1$ region, thus the $N$-baryon trajectory is the most 
useful for the extraction of the mass parameters and for the validation of this  model.

A few words about the precision on the extracted masses are
necessary. Table~\ref{table} shows the fit values with the necessary  {\em numerical}  precision  
to reproduce the $\chi^2/ndf$ and thus the proton mass, 
which has a very small experimental
uncertainty. 
The uncertainties on the
extracted parameters from the fits are negligible, therefore, 
they are not quoted (see Appendix~A for details).
We did not estimate the exact range of the fit parameters 
which give an acceptable fit (i.e.  with  $\chi^2/ndf < 1$) since this
will require a significant computational time. However,  from several tests
we did, we have concluded that the parameters are fairly close to 
$m_1\simeq 0.27\pm 0.01 \gev$,
$m_2\simeq 0.11\pm 0.01\gev$ and $\sigma \simeq 0.92\pm 0.1\gev^2$ (the so-called
Fit I), with a very
strong and complicated correlation between the values. 

In addition to the solution given above, we have found another solution
with  $m_1=0.228\gev$,
$m_2=0.179\gev$ and $\sigma=0.920\gev^2$ (Fit II, see Table C-1 of Appendix~C), 
which also gives
an acceptable fit, $\chi^2/ndf =0.9/1$. 
In this case, both masses are rather similar, almost in the spirit of the
expectations discussed in \cite{selem}.  
This solution corresponds to
a different radius of the quark-diquark system (see a discussion in 
Appendix~C).         
We would like to note
that the obtained parameters may not be the only solutions which
lead an acceptable  fit with  $\chi^2/ndf < 1$, thus
the {\em model} uncertainties are unknown yet.  
To find possible alternative solutions will require a 
significant computational time and, therefore, 
this is outside of the scope of this paper.
The model uncertainties for the $N$-baryon trajectory are further discussed 
in Appendix~C.

It was also verified that if one  sets  $m_1=m_2$ for the fit using the numerical 
solution, then the fit cannot be considered as a good one, $\chi^2/ndf\simeq 5/2$.
Obviously, the assumption $m_1=m_2$ is more appropriate for mesons, and it
is remarkable that the $N$-baryon trajectory does not support it.

%%%%%%%%%%%%%%%%%%%%%%%%%%%%%%%%%%%%%%%%%%%%%%%%%%%%%%%%%%%%%%%%%%%%%%%
\section{Mesons}
\label{sec4}

It would be interesting  to see 
the applicability of Eq.~(\ref{eq2}) and the assumptions discussed in Sect.~\ref{sec1}
to the mesonic  sector.  
We will mostly be interested in 
the linearity of the trajectory which bounds all known  mesons on the $(J,M^2)$ plane
at low masses. This is very similar to the $\Delta$ baryons
which bound the entire baryonic sector
at low masses by the straight trajectory shown in Fig.~\ref{fig2delta+}.    

Figure~\ref{fig3} shows the $(J,M^2)$ plane for all PDG mesons (filled symbols).
As before, the filled circles
show all mesons with lowest masses (for $J\geq 1$) which were
used in the linear fit.
It can be seen that such mesons approximately lie on a straight line, and this  
can scarcely be a coincidence.
The fit has a significant $\chi^2/ndf$ due to very small experimental 
uncertainties on the measured meson masses, thus  
one can judge  about the linearity only with a certain caution. 
There is only one
significant residual at $J=5$; however, it is not improbable  that
the $X(2210)$ state may  have $J=5$ and thus it falls exactly on the
linear fit. All other states indicated with the open circles
have unknown  $J$ and their locations were hypothesized.

While the slope for the meson sector is somewhat lower, it should be noted
that the
upper $95\%$ confidence-level line ($p_0= -0.503, p_1= 1.099$)
approximately coincides with that for the baryonic  sector.

It is interesting to note that there is a clear periodicity  
in space and charge conjugation parity indicated  as ($P,C$)  
for lowest-mass mesons which bound the meson spectrum on the $(J,M^2)$ plane
(see Fig.~\ref{fig3}). This was already noted for the trajectory of vector mesons in \cite{solo}.       
Based on this observation, we can predict that the $C$-parity of $X(2750)$ is likely to be negative. 

\section{Conclusion}

In this paper, using the most recent experimental data,  
empirical rules have been proposed which are sufficient to reveal      
a strict linear relationship between
the total angular momentum and squared-mass
of baryons. 
Using a least-squares regression, we did not find
evidence for non-linearity of such trajectories even when 
the most stringent statistical tests based on the $\chi^2$ and $95\%$ confidence-level intervals 
were  used. 
In  one case, this conclusion was made after the analysis of a number of baryons
as large as six.
This  observation may provide the basis for a new systematization
of hadrons and
certainly require thoughtful theoretical investigation.

The linear $N$-baryon trajectory  can be well described by 
a semi-classical approach based on 
a spinning relativistic
string with diquark and quark at the ends.
We have shown this baryonic trajectory
is $\chi^2$ consistent with the exact numerical solution
of this model (up to nine digits of the extracted model parameters).  
In this paper we have determined the (di)quark masses from  
the fit of the  $N$-baryon trajectory using our
solution.

In principle, this semi-classical model is qualitatively  consistent
with other linear baryonic trajectories. However, in this paper, we did not attempt to     
verify this using our numerical calculations.  
Also, we  did not analyse mesons as carefully as baryons. However,   
it is interesting to observe that
the meson spectrum on the $(J,M^2)$ plane is restricted at small masses by  a linear trajectory formed
by mesons with $S=C=B=0$ which have
periodicity
in space and charge conjugation parity.

\section*{Acknowledgements}
\vspace{0.3cm}
We thank Prof.~E.~Lohrmann 
for discussion of the results of this paper and 
A.~Selem for correspondence and an explanation of some results
of Ref.~\cite{selem}.

\newpage
\appendix
\renewcommand{\theequation}{A-\arabic{equation}}
% redefine the command that creates the equation no.
\setcounter{equation}{0}  % reset counter
\section*{APPENDIX~A}  % use *-form to suppress numbering
\label{appA}

The linear regression was performed using the weights
$w=1/\sigma^2$, where  $\sigma$
is the experimental uncertainty on the mass squared of a hadron.
% Such uncertainty is  related to a hadron mass $M$ and its experimentally
% measured uncertainty $\delta M$ as $\sigma=2 M \delta M$.
In case of asymmetric experimental errors, we take the average of the upper and the
lower experimental uncertainty.

The least-square linear regression was carried out using the R program
\cite{RR}. 
The R-package was also used to estimate confidence-level intervals  on the linear regression.
This package gives larger uncertainties on the linear-regression
parameters than the MINUIT program from the CERNLIB FORTRAN library~\cite{cernlib}.   
% As example, a linear fit performed by MINUIT for the $N$-baryon trajectory 
% shown in Fig.~\ref{fig2n+} gives:
% $p_0=0.393343 \pm 0.000002$ and $p_1= 0.974022 \pm 0.000005$ 
% (see Table~\ref{table_fit} for comparison).  
The MINUIT parameter errors give information   on the uncertainty in the best
fit values and are not meaningful when
points do not have (or have very small) experimental uncertainties.

In contrast to the standard linear least-square regression analysis,
the MINUIT program was used for validation of the relativistic string model discussed
in Section~\ref{sec3}.
This simplifies the fit procedure in case of non-linear functions,
especially when their analytic form is unknown (as in case
of the numerical solution of the string model discussed in Section~\ref{sec3}). 
The extracted fit parameters with the necessary
numerical precision to reproduce the proton mass are shown in Table~\ref{table}. 
The MINUIT parameter errors are not shown, since they are smaller than the last digit.

% \newpage
% redefine the command that creates the equation no.
\renewcommand{\theequation}{B-\arabic{equation}}
\renewcommand{\thefigure}{B-\arabic{figure}}
\renewcommand{\thetable}{B-\arabic{table}}
\setcounter{equation}{0}  % reset counter 
\section*{APPENDIX~B}  % use *-form to suppress numbering
\label{appB}

Here we will derive the functions used 
to fit the $N$-baryon trajectory shown  
in Fig.~\ref{a_boris_n+}. 
First, recall the main set of equations (8)-(9) of the string model \cite{selem}:
\begin{eqnarray}
E&=&\sum_{n=1}^{2} \Big (m_n\gamma_n +\frac{T}{\omega}\arcsin(\beta_n)\Big ), \label{ap_1}\\
J&=&\sum_{n=1}^{2} \Big [\frac{m_n}{\omega}\beta^2_n \gamma_n
+\frac{T}{2\omega^2} \Big ( \arcsin (\beta_n)-\frac{\beta_n}{\gamma_n}\Big )\Big ], \label{ap_2}\\
T&=&\omega m_n \beta_n \gamma^2_n,\ \ \ \ n=1,2. 
\label{ap_3}
\end{eqnarray}
Here, $E$ is the energy of the quark-diquark system rotating with 
an angular momentum $J$, 
$m_n$ denotes the mass of a diquark ($n=1$) and quark ($n=2$), $\omega$ is angular velocity, $T$ is 
a string tension and  
$\beta_n$ is a linear velocity. The factor $\gamma_n$ of 
a given (di)quark is defined as
$$
\gamma_n=\frac{1}{\sqrt{1-\omega^2 r^2_n}},
$$
where $r_n$ is a  distance from the center of rotation.

Our goal is to find a relationship between the energy $E$ and $J$.
Such a relationship has to be  
expressed in terms of $\sigma=2\pi T$ and $m_n$. 
Excluding 
terms with  $\arcsin(\beta_n)$ from 
Eq.~(\ref{ap_1}) and  
using the representation 
$$
\arcsin(\beta_n)=\frac{\pi}{2}-\frac{1}{2}\arcsin \Big (2\frac{\beta_n}{\gamma_n}\Big ),
$$
one gets from Eq.~(\ref{ap_1})-(\ref{ap_3}) 
\begin{eqnarray}
E&=&2\omega J + \frac{m_1}{\gamma_1}+ \frac{m_2}{\gamma_2} \label{ap_4},\\
J&=&\frac{1}{2T}\Big (\frac{T}{\omega}\Big )^2\Big [\pi 
+\frac{1}{2}\sum_{n=1}^{2} \Big (\frac{2\beta_n}{\gamma_n} 
- \arcsin\Big (\frac{2\beta_n}{\gamma_n}\Big ) \Big ) \Big ]
\label{ap_5},\\
x_n&=&\frac{1}{\beta_n\gamma^2_n},\ \ \ \ n=1,2. 
\label{ap_6}
\end{eqnarray}
We use  Eqs.~(\ref{ap_4})-(\ref{ap_6}) to fit the $N$-baryon trajectory and to extract
the parameters $m_n$ and $\sigma$ from the data. 
At fixed $J$, $m_n$ and $\sigma$, the  equation (\ref{ap_5}) can be solved 
relative to   $\omega$   either numerically, or to some approximation,  using a series 
expansion. Here, $x_n= m_n\omega/T$ can be used 
as the expansion variable in the limit of small (di)quark masses.

In terms of $x_n$,
\begin{eqnarray}
\frac{1}{\gamma_n}&=&\Big [ \frac{2x_n}{x_n+\sqrt{4+x^2_n}}\Big ]^{1/2}= x^{1/2}_n
-\frac{1}{2^2}x^{3/2}_n +\frac{1}{2^5}x^{5/2}_n    + {\cal{O}} (x^{7/2}_n) \label{ap_7},  \\
\frac{\beta_n}{\gamma_n}&=&\frac{1}{x_n}\Big [\frac{2x_n}{x_n+\sqrt{4+x^2_n}} \Big ]^{3/2}
= x^{1/2}_n  -\frac{3}{2^2}x^{3/2}_n + \frac{9}{2^5}x^{5/2}_n
+ {\cal{O}} (x^{7/2}_n).
\label{ap_8}
\end{eqnarray}
Using the expansion 
$$\arcsin \Big (2\frac{\beta_n}{\gamma_n}\Big )=
\frac{2\beta_n}{\gamma_n}+\frac{4}{3} \Big (\frac{\beta_n}{\gamma_n}\Big )^3 
+\frac{12}{5} \Big (\frac{\beta_n}{\gamma_n}\Big )^5
+ {\cal{O}} (\Big (\frac{\beta_n}{\gamma_n}\Big )^7),$$
Eq.~(\ref{ap_5}) becomes
\beq
J=\frac{1}{2T}\Big (\frac{T}{\omega}\Big )^2\Big [\pi 
-\frac{2}{3} \Big (\frac{\beta_1}{\gamma_1}\Big )^3 
-\frac{2}{3}\Big (\frac{\beta_2}{\gamma_2}\Big )^3 
-\frac{6}{5}\Big (\frac{\beta_1}{\gamma_1}\Big )^5 
-\frac{6}{5}\Big (\frac{\beta_2}{\gamma_2}\Big )^5 
+{\cal{O}} (\Big (\frac{\beta_n}{\gamma_n}\Big )^7)
\Big ].\label{ap_9}
\eeq
Thus, substituting Eq.~(\ref{ap_7}) and Eq.~(\ref{ap_8}) in
Eq.~(\ref{ap_4}) and Eq.~(\ref{ap_9}) and keeping only terms up to order $m^{5/2}_n$, one obtains: 
\begin{eqnarray}
E&=&\pi\Big (\frac{T}{\omega}\Big ) + \frac{\mu^{3/2}_1}{3}\Big ( \frac{\omega}{T}\Big )^{1/2} 
+ \frac{\mu^{5/2}_2}{20}\Big ( \frac{\omega}{T}\Big )^{3/2}
+ {\cal{O}} (\mu^{7/2}), \label{ap_10}\\
J&=&\frac{\pi}{2T}\Big (\frac{T}{\omega}\Big )^2 - \frac{\mu^{3/2}_1}{3T}\Big (\frac{\omega}{T} \Big )^{-1/2}
+\frac{3}{20}\frac{\mu^{5/2}_2}{T}\Big (\frac{\omega}{T} \Big )^{1/2}
+ {\cal{O}} (\mu^{7/2}) 
\label{ap_11}
\end{eqnarray}
with $\mu^{3/2}_1=m^{3/2}_1+m^{3/2}_2$ and $\mu^{5/2}_2=m^{5/2}_1+m^{5/2}_2$.

Let us denote
$V=\sqrt{\omega / T}$
and rewrite Eq.~(\ref{ap_11}) as a quartic equation
\beq
V^4 +aV^3 -b = 0,
\label{ap_12}
\eeq
with
$$a=\frac{2\pi\mu^{3/2}_1}{3\sigma J},\ \ \ \ \ \ b=\frac{\pi^2}{\sigma J},$$
where the term proportional to $\mu^{5/2}_2$ was neglected.
To solve the quartic equation (\ref{ap_12}) we use the Ferrari method \cite{mathword}.
The resolvent cubic  of Eq.~(\ref{ap_12})
\beq   
U^3 + 4bU + ba^2 = 0,
\label{ap_13} 
\eeq
transforms Eq.~(\ref{ap_12}) to a  quadratic equation: 
\beq
V^2 +\gamma_{\pm}V + \delta_{\pm}=0
\label{ap_14} 
\eeq
with
\beq
\gamma_{\pm}=\frac{1}{2}[a \pm \sqrt{a^2+4U}], \ \ \ 
\delta_{\pm}=\frac{1}{2}[U \pm\sqrt{U^2+4b}],
\eeq
where a real root of Eq.~(\ref{ap_13}) (the Cardano formula) is 
\beq
U= \frac{1}{6}\Big [12b\sqrt{81a^4+768b}-108ba^2\Big ]^{1/3}
 -\frac{1}{6}\Big [12b\sqrt{81a^4+768b}+108ba^2\Big ]^{1/3}.
\eeq
Thus, a real and positive solution of Eq.~(\ref{ap_14})  and therefore,  of 
Eq.~(\ref{ap_12}), is
\beq
V=\Big (\frac{\omega}{T} \Big )^{1/2}=\frac{1}{4}\Big \{\Big [(a^2+4U)^{1/2} - a \Big ]
+\sqrt{\Big [(a^2+4U)^{1/2} - a\Big ]^2   + 8\Big [(U^2+4b)^{1/2} - U\Big ]}\Big \}.
\label{ap_17}
\eeq

Substituting  the above expression in  Eq.~(\ref{ap_10}), one obtains a desired relationship
between  the baryonic mass $M=E$ and variables $J$, $\sigma$ and $m_n$.

% \appendix
\renewcommand{\theequation}{C-\arabic{equation}}
\renewcommand{\thefigure}{C-\arabic{figure}}
\renewcommand{\thetable}{C-\arabic{table}}
% redefine the command that creates the equation no.
\setcounter{equation}{0}  % reset counter
\setcounter{figure}{0}  % reset counter
\setcounter{table}{0}  % reset counter
\section*{APPENDIX~C}  % use *-form to suppress numbering
\label{appc}

Using  a graphical representation of different parts of 
Eqs.~(\ref{ap_4}) and (\ref{ap_5}),  it is easy
to demonstrate that the string model described  
by the system of equations  Eqs.~(\ref{ap_4})-(\ref{ap_6}) allows several solutions for
the (di)quark masses.  Such solutions have trajectories with very small (but different)
non-linear behaviour.

\begin{figure}[h]
%\begin{center}
\hspace*{-6mm}
\begin{minipage}[h]{.49\textwidth}
\vspace*{8mm}
\includegraphics[height=8.5cm,width=8.0cm]{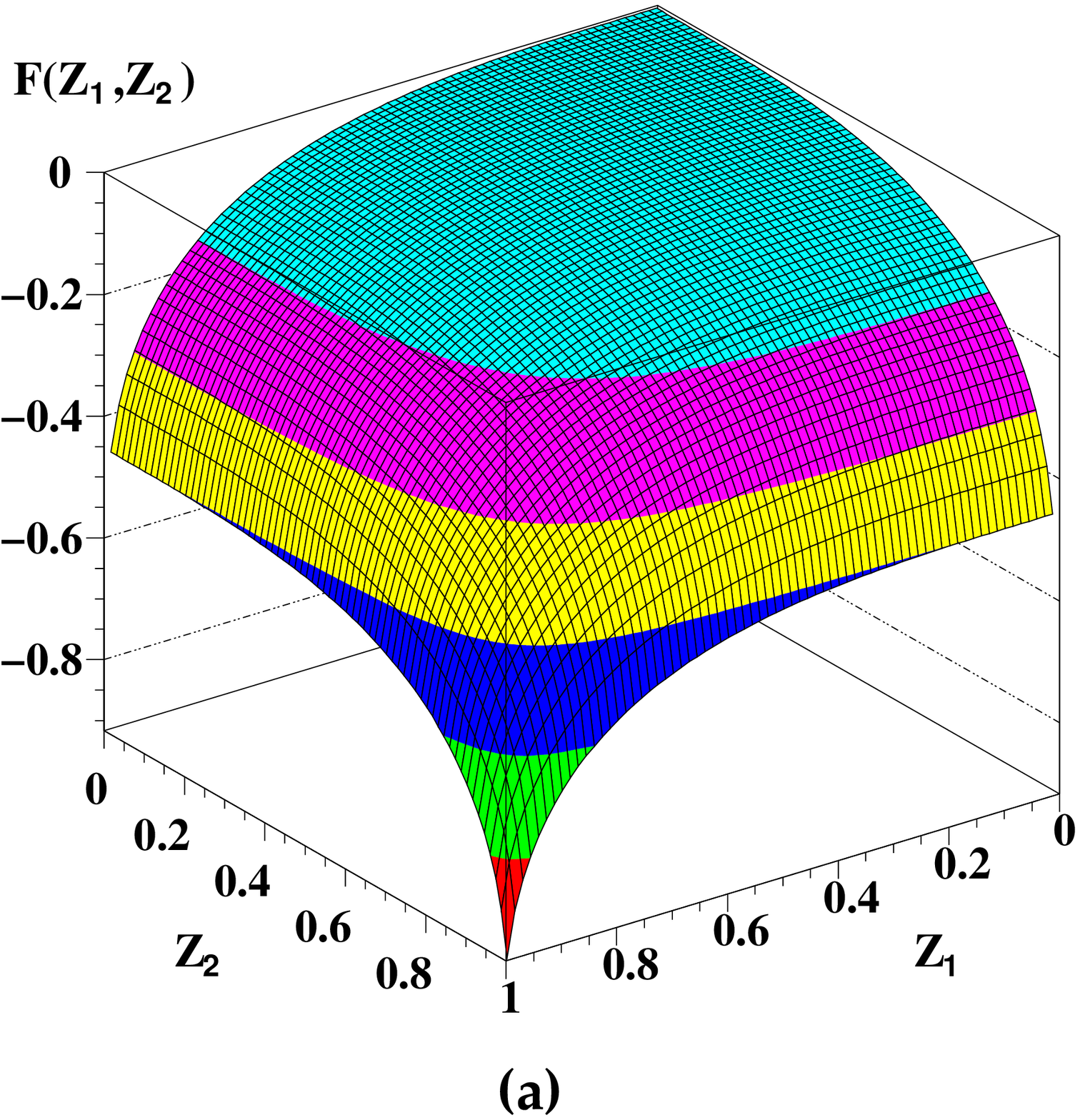}
\end{minipage}\hspace*{8mm}		
\begin{minipage}[h]{.49\textwidth}
\includegraphics[height=8.5cm,width=8.0cm]{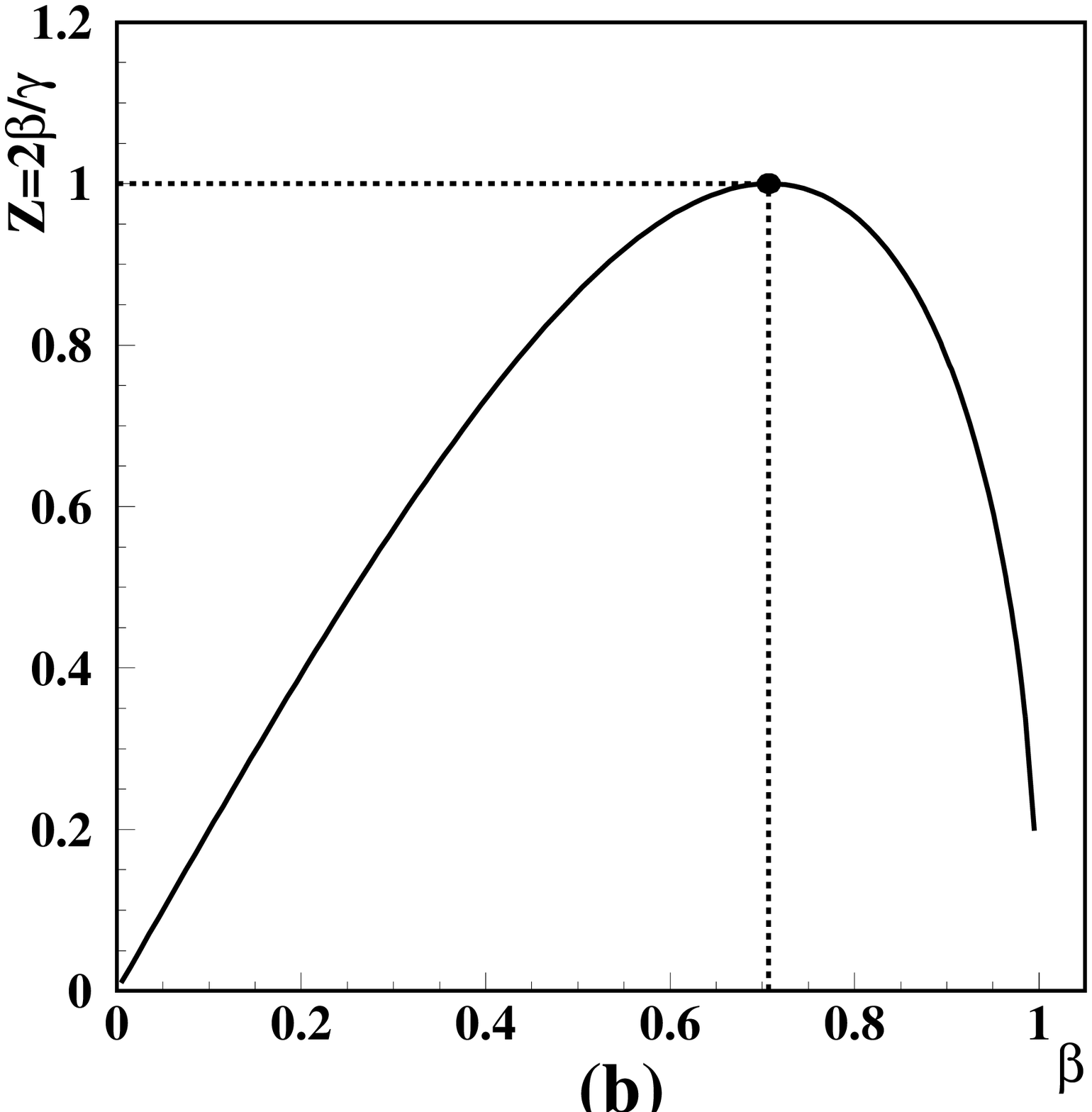}
\end{minipage}
\caption{$\F(\Za,\Zb)$ as a function of $\Za$ and $\Zb$ (left) and 
$\Zn $ shown as a function of the (di)quark velocity (right). }  
\label{APcc}
%\end{center} 
\end{figure}

Let us denote by $\Zn$ the following variables
\beq
\Zn = \frac{2\beta_n}{\gamma_n}, \ \ \ \  n=1,2
\eeq
and introduce a function
\beq
\F(\Za,\Zb)= \sum_{n=1}^{2} \Big (\Zn -\arcsin( \Zn )\Big )
\eeq
which represents the last term in Eq.~(\ref{ap_5}). The variables $\Zn$ have values in  the
interval $[0,1]$,  and are related to  the
parameters $m_n$, $\omega$ and $T$ through the equations Eq.~(\ref{ap_6})  and Eq.~(\ref{ap_8}).
The function $\F(\Za,\Zb)$ takes values in the range [$2-\pi,0$].
As shown in Fig.~\ref{APcc}(a), $\F(\Za,\Zb)$ is 
symmetric with respect to the median line $\Za=\Zb$ and a monotonic
function of the both arguments.
The main  feature of  Eq.~(\ref{ap_5})  is that 
the equation  $\F(\Za,\Zb)=\Fc$, where $\Fc$ is a constant,
defines an isoline on the surface and at fixed $J$ all the points $\{\Za,\Zb \}$
on this isoline correspond to the same angular velocity $\omega$. 
In  Fig.~\ref{APcc}(a), the isolines are shown as lines between colored  bands.
\vspace*{-3mm}
	
\vspace*{5mm}
Let us consider the mass evolution equation (\ref{ap_4})  starting from 
a very  high $J$.
As follows from (\ref{ap_5}) and the  definition of $x_n$,  
a very high $J$ can be obtained only at very small $\omega$ or  $x_n$.
Then Eqs. (B-7) and (B-8) tells that $\gamma_n \gg 1$, 
$\beta_n\approx 1$,  ${\Zn}\approx 0$ and $\F(\Za,\Zb)\approx 0$.
Thus, at such $J$, $\beta_n$ and $\omega$, the quark and di-quark 
are at large distances from the center of rotation.

Now, if  $J$  decreases and  $\omega$, consequently, increases,  
the phase point $(\Za,\Zb)$ on the surface $\F(\Za,\Zb)$ 
 is moving along some path in the direction of $(\Za=1,\Zb=1)$. 
A fit of the $N$-baryon trajectory  fixes 
a particular path passing at $J=1/2$ through a point $(\Za^{(f)},\Zb^{(f)})$
on the surface  given by $\F(\Za,\Zb)$. 
If  $\Zn $ function is  plotted as a function of the (di)quark velocity 
$\beta_n$, as shown in Fig.~\ref{APcc}(b), one finds that 
the point is moving from $\beta\approx 1$ to a smaller $\beta$. 
 The maximum $\Zn =1$ is reached at $\beta_0 =1/\sqrt{2}$.

\begin{table}
\begin{center} 
\begin{tabular}{|l|c|c|c|}
\hline
&Numerical solution&Numerical solution&limit $\Za=\Zb=1$,\\
Parameters&Fit I     & Fit II     &   \\
\hline
$\chi^2/ndf$&0.9/1&0.9/1& - \\
$m_1$, GeV& 0.271713&0.228808& 0.18393 \\
$m_2$, GeV& 0.114134& 0.179558& 0.18393 \\
$\sigma$, GeV$^2$& 0.923374& 0.920000& 1.12428\\
\hline
$J$&1/2&1/2&1/2\\
$\omega$, GeV& 0.643476& 0.628674& 0.679332 \\
$\Za^{(f)}$& 0.935557& 0.974606&1\\
$\Zb^{(f)}$& 0.975607& 0.998309&1\\
$\F(\Za^{(f)},\Zb^{(f)})$& -0.648139& -0.884682& 2$-\pi$\\
$x_1$ & 1.189721& 0.982401&1/$\sqrt{2}$\\
$x_2$ & 0.499746& 0.770943&1/$\sqrt{2}$\\
$\beta_1$ & 0.822549& 0.782280&1/$\sqrt{2}$\\
$\beta_2$ & 0.780872& 0.727369&1/$\sqrt{2}$\\
$r_1$, GeV$^{-1}$ & 1.278291& 1.244335& 1.045218\\
$r_2$, GeV$^{-1}$ & 1.213521& 1.156988& 1.045218\\

\hline
\end{tabular}
\end{center}
\label{tableC}
\caption{The model parameters extracted from Eqs.~(\ref{ap_6})-(\ref{ap_8}).  }

\end{table}

Table~C-1 gives the model parameters\footnote{We have reduced the numerical precision
of the quoted fit values for representative purposes. 
The exact fit values can be found from Table~2.} extracted from  
Eqs.~(\ref{ap_6})-(\ref{ap_8}) using  
$m_n$ and  $\sigma$ found from Fit~I and Fit~II.
First of all, let us note that 
all $\beta_n <1/\sqrt{2}$
(or higher values of $\omega$) correspond to a non-physical $J<1/2$ and
$\beta_n  > 1/\sqrt{2} $ holds for the both fits.

The baryon masses predicted by the string model  are degenerated 
with respect to  $\omega$ and $m_n$. Indeed,  $\F(\Za,\Zb)$ is negative and, 
therefore,  there is a range of $\omega$ for which   
one obtains $J_1 = J_2$ at different $\omega_1$ and $\omega_2$. 
If (B-5) is equated at different $\omega$, then
\beq
\frac{1}{\omega^2_1}[2\pi+\Fc_1] = 
\frac{1}{\omega^2_2}[2\pi+\Fc_2]
\eeq 
or
\beq
\Fc_2= \Big (\frac{\omega_2}{\omega_1}\Big )^2(2\pi+\Fc_1) -2\pi . 
\eeq 
Here $\Fc_1$ and $\Fc_2$ denotes $\F(\Za,\Zb)$ at two phase-space points
$(\Za,\Zb)_{1,2}$. The condition
\beq
2-\pi \leq \Fc_2\leq 0
\eeq 
puts the following restriction on $\omega_2$ (at a fixed $\omega_1$): 
\beq
\frac{2+\pi}{2\pi+\Fc_1}\leq  \Big (\frac{\omega_2}{\omega_1}\Big)^2
\leq\frac{2\pi}{2\pi+\Fc_1} .
\label{C6}
\eeq 
The condition (\ref{C6}) could mean that
if one finds a set of parameters $\{m_1,\, m_2,\, \sigma\}_{_1}$ with the  evolution
 along a path  $\{\omega,\, \beta_1,\, \beta_2\}_{_1}$, there should
 exist another solution $\{m_1,\, m_2,\, \sigma\}_{_2}$ with a different 
 path
$\{\omega,\, \beta_1,\, \beta_2\}_{_2}$. The second found solution 
for the $N$-baryon trajectory may confirm such interpretation.
Thus, we conclude that the model should have an intrinsic uncertainty 
for the extracted masses depending on the $\Za$ and $\Zb$ values.

%%%%%%%%%%%%%%%%%%%%%% references %%%%%%%%%%%%%%%%%%%%%%%%%%%%%%
% \bibliographystyle{./Macros/h-elsevier-clean}
\bibliographystyle{./Macros/l4z_pl}
\def\bibname{\Large\bf References}
\def\refname{\Large\bf References}
\pagestyle{plain}
\bibliography{hadrons}

%%%%%%%%%%%%%%%%%%%%%%%%%%%%%% FIGURES %%%%%%%%%%%%%%%%%%%%%%%%%%%%%%

\newpage
\renewcommand{\theequation}{\arabic{equation}}
\renewcommand{\thefigure}{\arabic{figure}}
\renewcommand{\thetable}{\arabic{table}}
\setcounter{equation}{0}  % reset counter
\setcounter{figure}{0}  % reset counter
\setcounter{table}{0}  % reset counte

\begin{table}
\centering
\vspace{2.0cm}
\begin{tabular}{|l|r|r|r|}
 \hline
Particle family & $p_0$, GeV$^2$  & $p_1$, GeV$^2$   & $\chi^2/ndf$   \\
 \hline
$N$ baryons, $P=+1$  & $0.393\pm 0.003$  & $0.974\pm 0.006$ & 0.5/2  \\ 
$\Delta$ baryons, $P=+1$  & $-0.156\pm 0.013$  & $1.116\pm 0.006$ & 6.1/4  \\
$\Lambda$ baryons, $P=-1$  & $0.746\pm 0.007$  & $1.042\pm 0.004$ & 0.2/1  \\
$\Sigma$ baryons, $P=+1$  & $0.277\pm 0.024$  & $1.098\pm 0.016$ & - \\
mesons (lowest mass) & $-0.454\pm 0.026$  & $1.052\pm 0.019$  & 220/5  \\ 
\hline
\end{tabular}
\caption{
Fit parameters obtained using the linear parameterisation $M^2 = p_0 + p_1 J$  for several 
hadronic families with the lowest masses on the $(J,M^2)$ plane. The $\chi^2/ndf$ for the $\Sigma$
states is not given since there are only two points for the linear fit (see the text).  
}
\label{table_fit}

\end{table}

\begin{table}
\centering
\begin{tabular}{|l|c|c|c|c|}
 \hline
 Parameters      & SW fit     & ${\cal{O}} (\mu^{3/2})$ & ${\cal{O}}
(\mu^{5/2})$ & numerical \\
                 &            & solution               & solution
& solution   \\
 \hline
$\sigma$, GeV$^2$& 0.908091023 & 0.889953024            & 0.902776942 
& 0.923374256   \\
$\mu$, GeV       & 0.323158469 & -                     & -
&  -  \\
$m_1$, GeV       &  -          & 0.302678728           & 0.280596424 
& 0.271713077  \\
$m_2$, GeV       &  -          & 0.109731744           & 0.109180448 
& 0.114133666   \\
$\chi^2/ndf $    & 1.5/2       &  2.2/1                & 1.8/1
& 0.9/1  \\
 \hline
 \hline
M$_p$, GeV &&&&\\
from Eq.~(\ref{ap_4})  & 0.938272037 & 0.938272031 & 0.938272034 & 0.938272048\\
 \hline
M$_p$, GeV  \cite{pdg} &\multicolumn{3}{c}{   $0.938272029 \pm 0.000000080$ 
} &\\
\hline
\end{tabular}
\caption{
Fit parameters obtained using different solutions of the relativistic
diquark model (see the text). The $N$-baryon trajectory was used for the
fit.
The parameter uncertainties are not shown (see 
Appendix~A).
The large number of digits, which is necessary to reproduce the quoted
$\chi^2/ndf$, is due to
a large experimental precision on the $N$-baryon masses.
The proton mass was calculated  from  Eq.~(\ref{ap_4}) using  different
approximations
and compared with the world average value.
}
\label{table}
\end{table}

\newpage
\begin{figure}
\begin{center}
\mbox{\epsfig{file=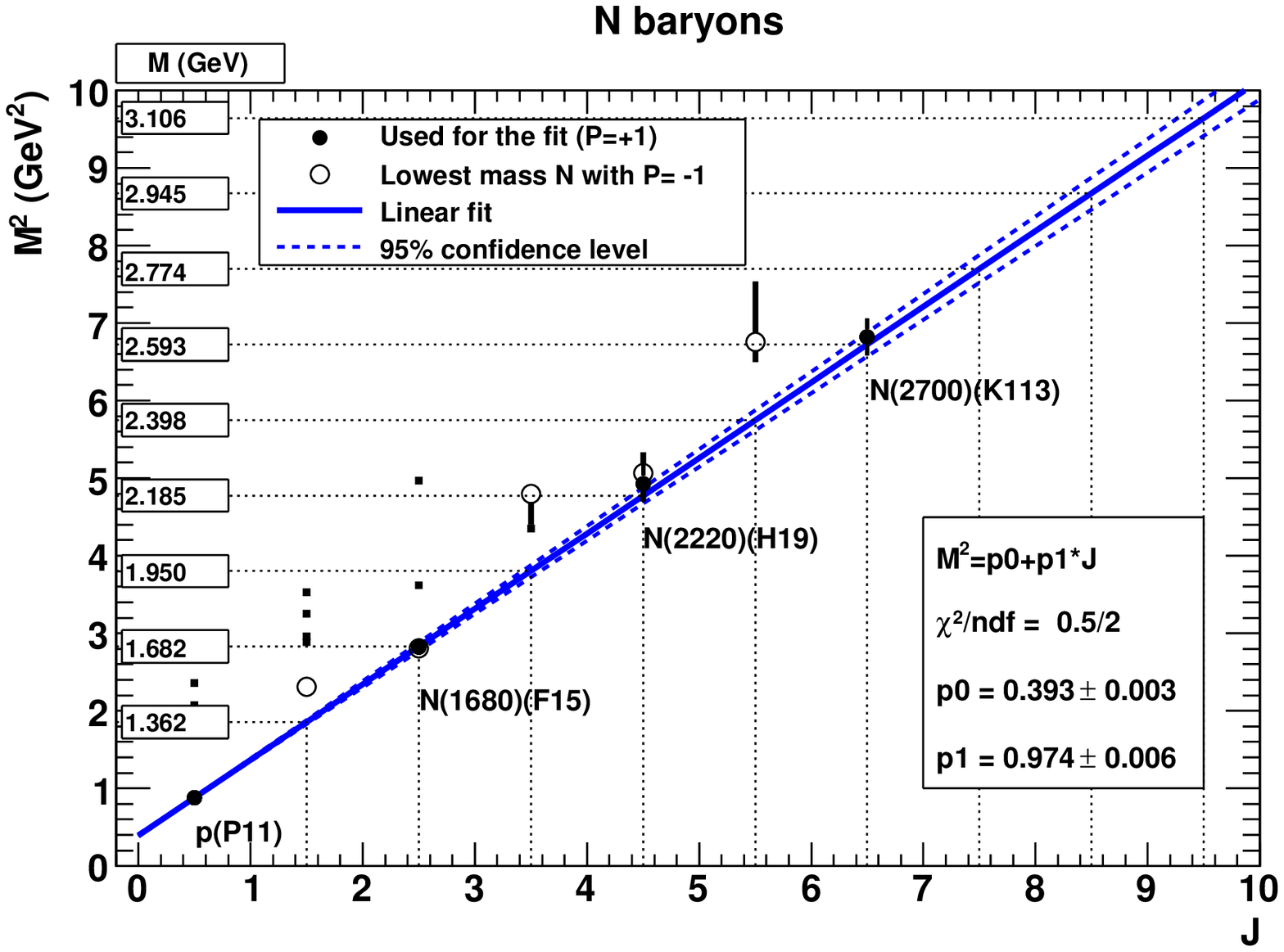,height=11cm}}
\caption
{The $(J,M^2)$ plane for $N$-baryons. The linear fit was
performed using the $P=+1$ baryons with smallest masses for a given $J$. 
The solid line shows the weighted-least squares regression
using a linear model and the dashed lines show the corresponding $95\%$ confidence-level interval.
The open symbols show lowest-mass baryons with $P=-1$: $N(1520)(D13)$ (with $J=3/2$),
$N(1675)(D15)$ (with $J=5/2$), $N(2190)(G17)$ (with $J=7/2$),  $N(2250)(G19)$ (with $J=9/2$) 
and $N(2600)(I111)$ (with $J=13/2$). The $N(1675)(D15)$ 
state overlaps in mass with the $N(1680)(F15)$ baryon with positive parity.     
}
\label{fig2n+}
\end{center}
\end{figure}

\newpage
\begin{figure}
\begin{center}
\mbox{\epsfig{file=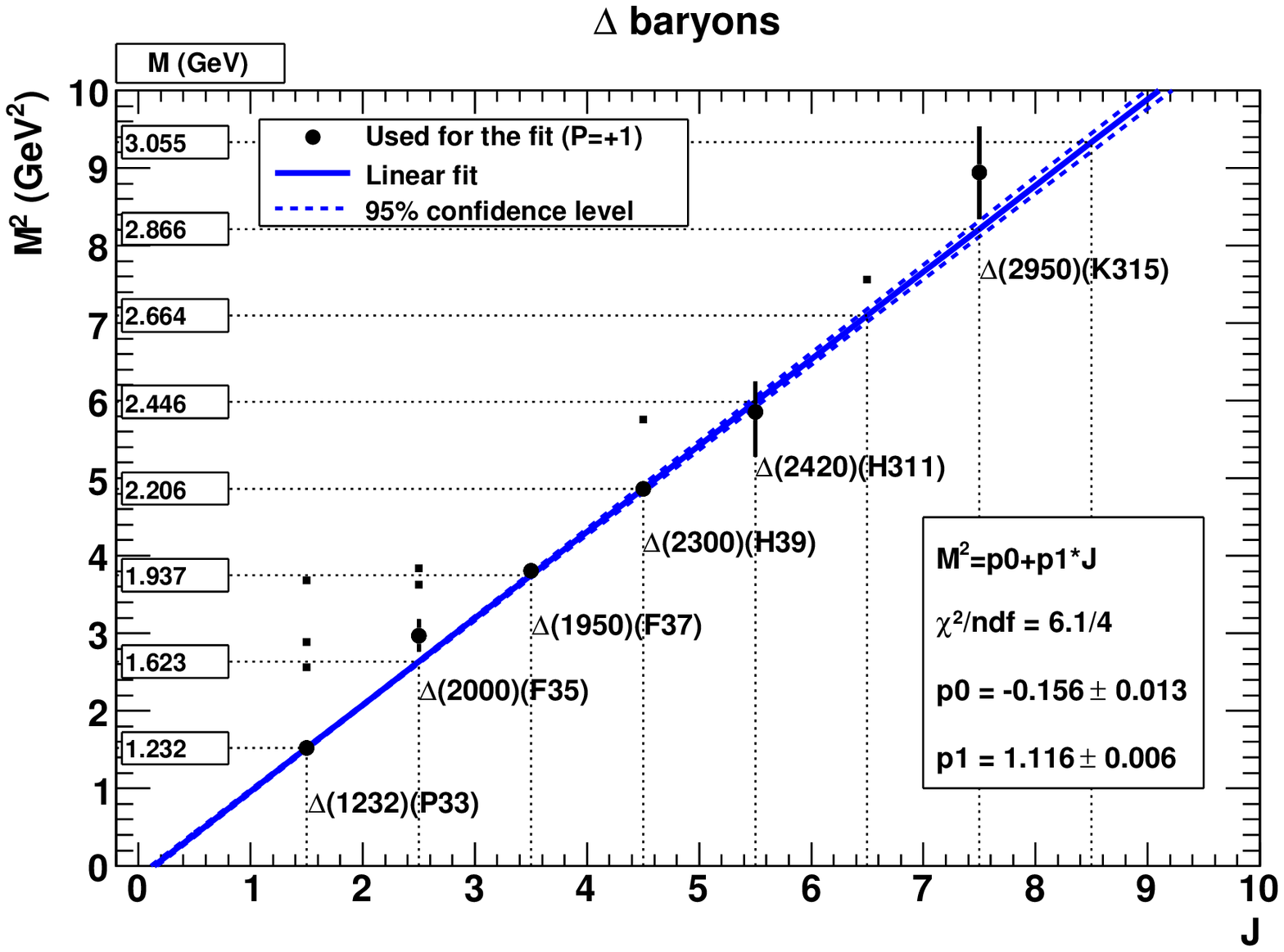,height=11cm}}
\caption
{
The ($J,M^2$) plane for $\Delta$ baryons with $P=+1$. The filled 
circles show lowest-mass $\Delta$ baryons  
used for the linear weighted-least squares regression. 
The solid line shows the fit, while the dashed line indicates
a $95\%$ confidence-level interval.
Small filled squares show all other $\Delta$  
baryons.
}
\label{fig2delta+}
\end{center}
\end{figure}

\newpage
\begin{figure}
\begin{center}
\mbox{\epsfig{file=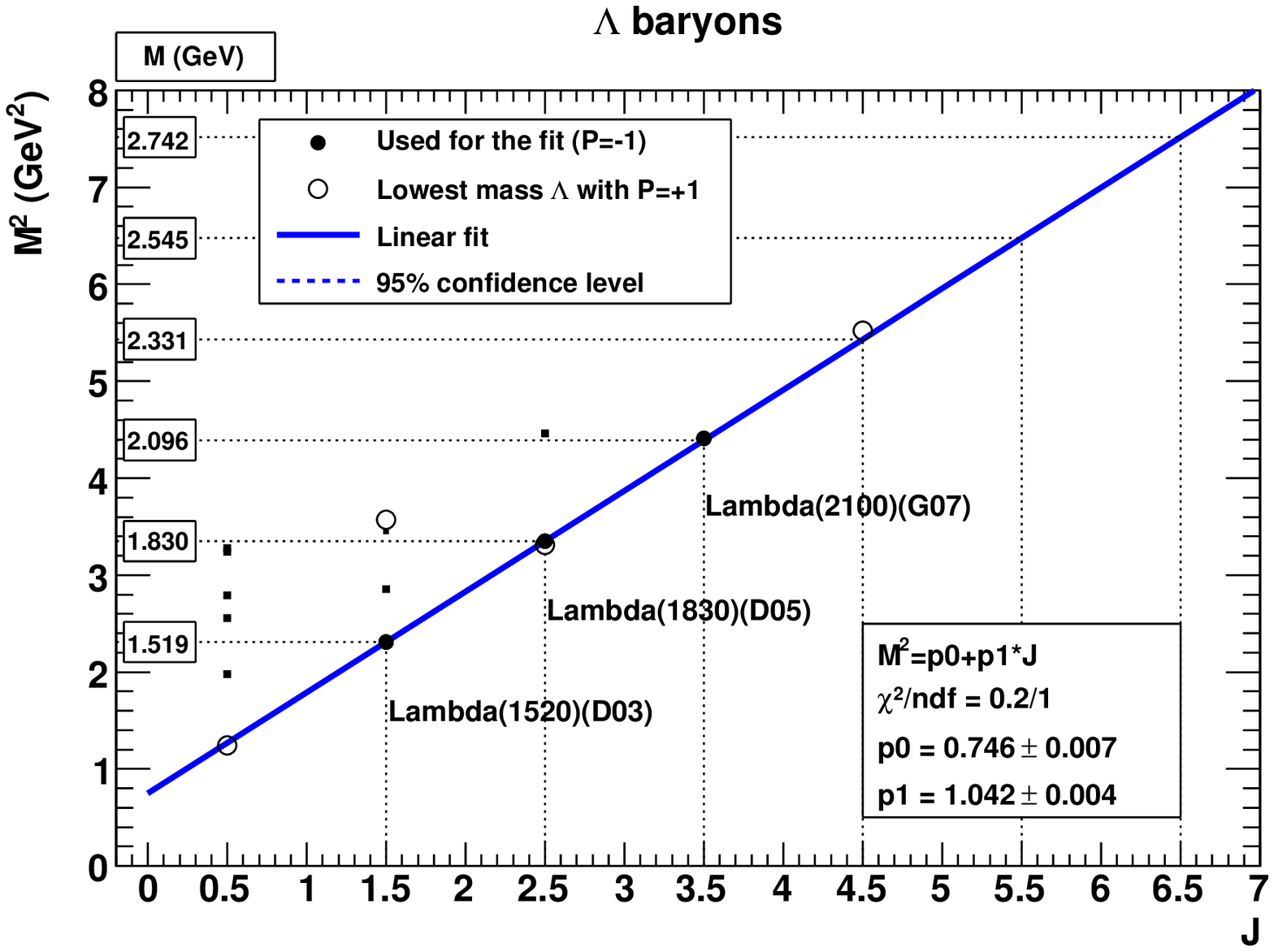,height=11cm}}
\caption
{
The ($J,M^2$) plane for $\Lambda$ baryons with $P=-1$ (filled symbols) and
$P=+1$ (open symbols). The latter hadrons correspond to
$\Lambda(1115)(P01)$, $\Lambda(1820)(F05)$, $\Lambda(1890)(P03)$ and $\Lambda(2350)(H09)$. 
The $\Lambda(1820)(F05)$ baryon with $P=+1$ is on top of $\Lambda(1830)(D05)$ with $P=-1$. 
Small filled squares show all other PDG
$\Lambda$ baryons.
}
\label{fig2lambda}
\end{center}
\end{figure}

\newpage
\begin{figure}
\begin{center}
\mbox{\epsfig{file=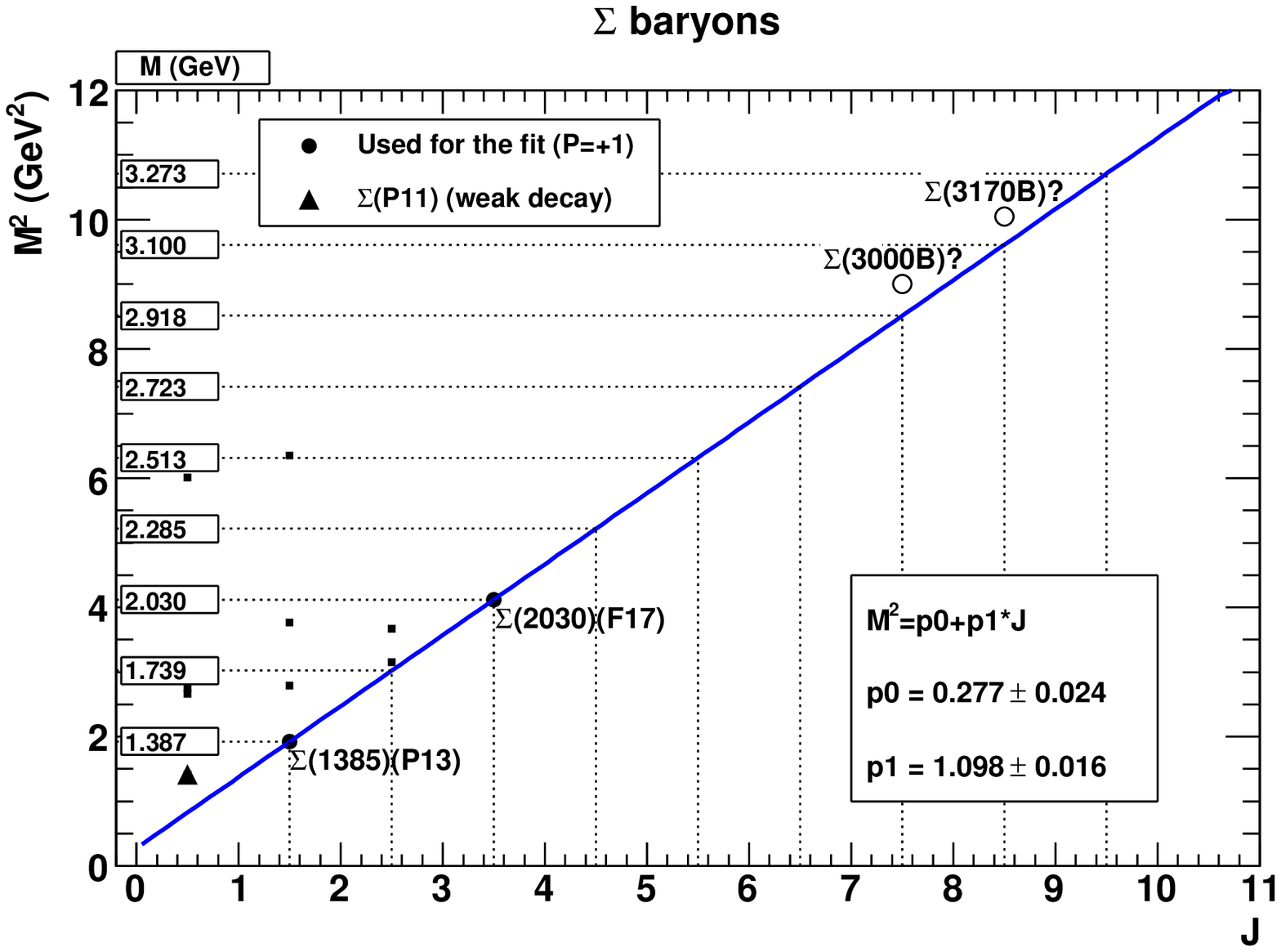,height=11cm}}
\caption
{
The ($J,M^2$) plane for $\Sigma$ baryons. The filled circles  
show the $\Sigma$ baryons with $P=+1$. 
The filled triangle shows the lowest-mass weakly-decaying
$\Sigma$ with $P=+1$, while the small filled squares show all other PDG $\Sigma$ baryons. 
The open circles indicate possible locations 
of several high-mass PDG baryons with unknown  $J$.
}
\label{fig2sigma+}
\end{center}
\end{figure}

\newpage
\begin{figure}
\begin{center}
\mbox{\epsfig{file=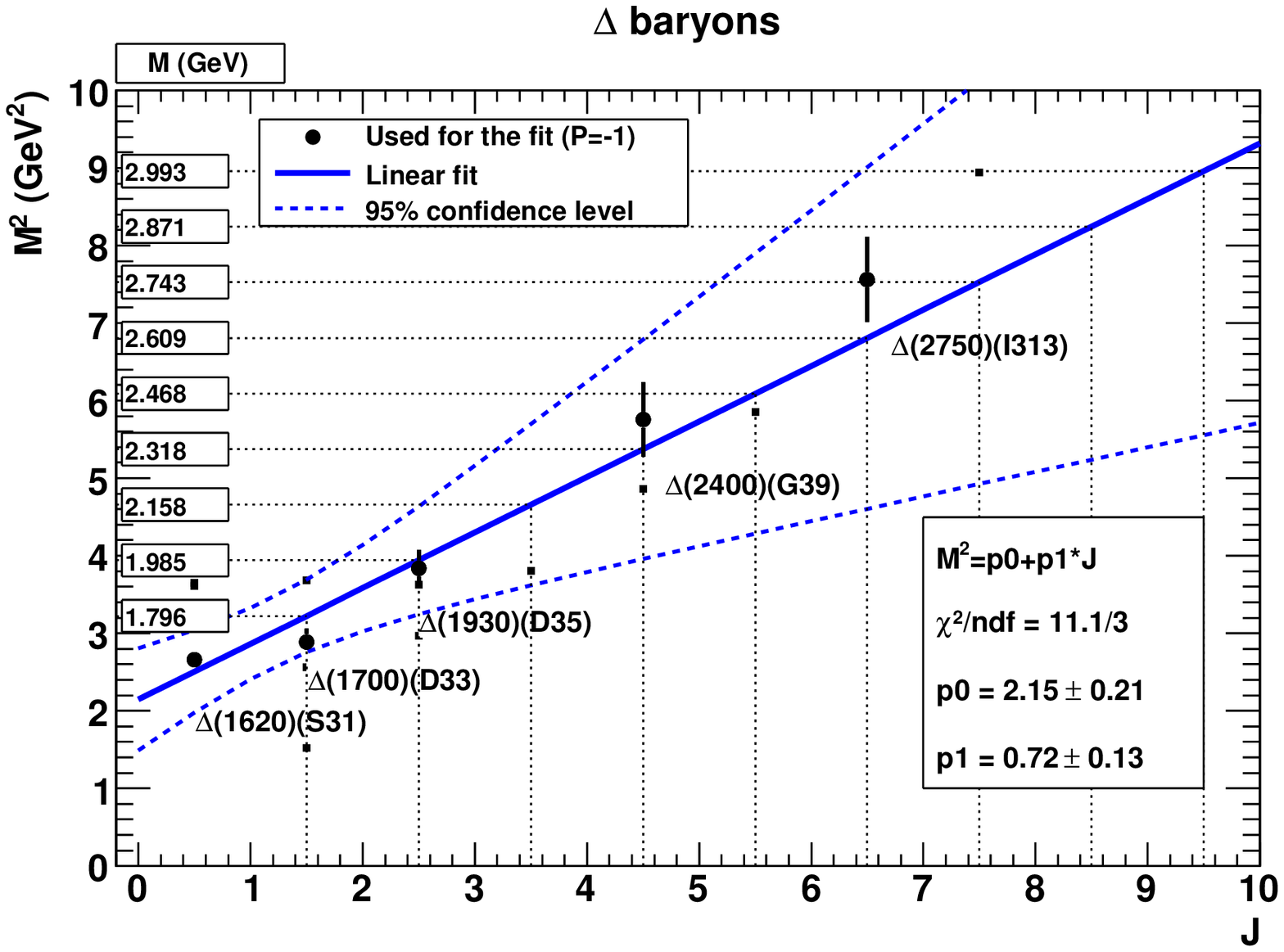,height=11cm}}
\caption
{
As Fig.~\ref{fig2delta+}, but the fit was performed for the 
lowest-mass $\Delta$ baryons with $P=-1$.
}
\label{fig2delta-}
\end{center}
\end{figure}

\begin{figure}
\begin{center}
\mbox{\epsfig{file=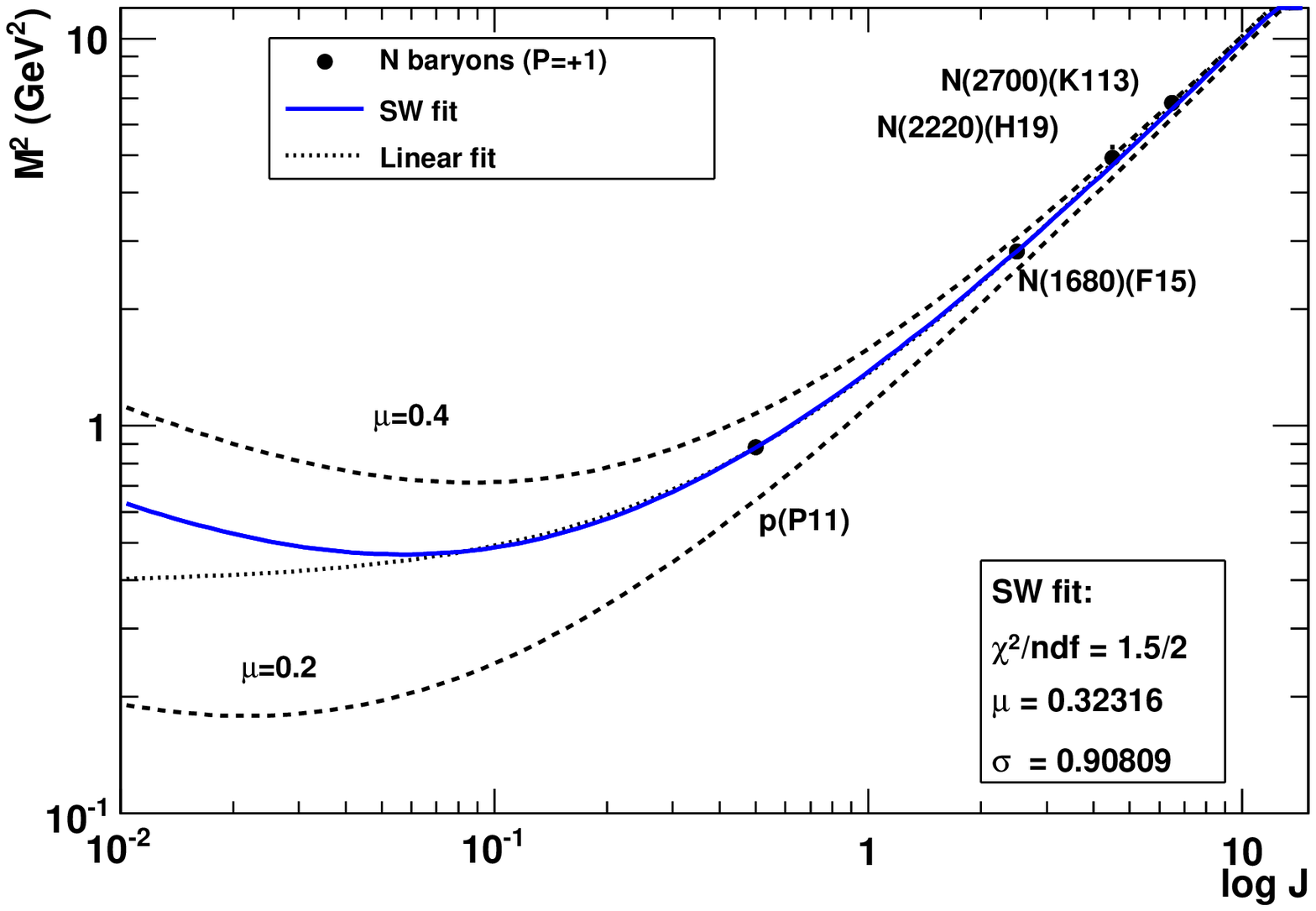,height=11cm}}
\caption
{The fit results for leading and daughter nucleons with positive parity.
Only baryons with the minimum mass requirement for a given $J$ were used.
The solid line indicated the fit by using Eq.~(\ref{sw1}) (the  so-called 
SW fit) with two free parameters, $\mu$ and $\sigma$. 
The parameter uncertainties 
are negligible (not shown). 
In order to illustrate the fit sensitivity to $\mu$,  the 
dashed lines show the same fit function but with
different $\mu$. The dotted line shows the linear fit shown in Fig.~\ref{fig2n+}.
}  
\label{a_selem_n+}
\end{center}
\end{figure}

\begin{figure}
\begin{center}
\mbox{\epsfig{file=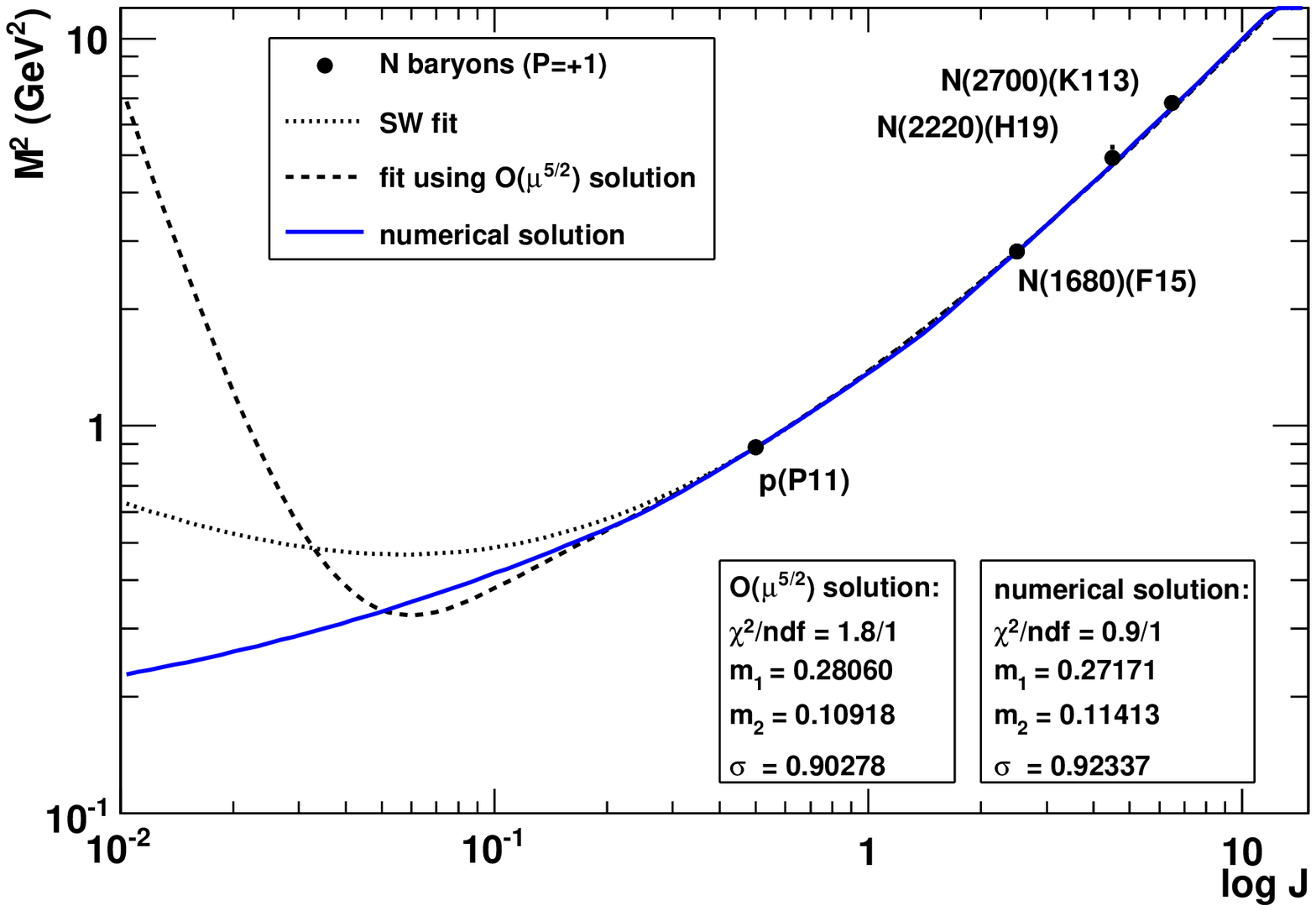,height=11cm}}
\caption
{
The dotted line shows the SW fit as in Figure~\ref{a_selem_n+}.
The dashed  line shows the  fit  using the solution
of the diquark string model  
calculated up to ${\cal{O}}  (\mu^{5/2})$ terms, see Eq.~(\ref{ap_10}) 
of Appendix~B.  
The fit was performed using two mass parameters,  $m_1$ and $m_2$ ($m_1>m_2$) and 
a string tension $\sigma$.
The solid line shows the fit using the exact numerical solution
(see the text). 
The parameter uncertainties
are negligible.
Table~\ref{table} gives more exact numbers from the fits.  
}
\label{a_boris_n+}
\end{center}
\end{figure}

\begin{figure}
\begin{center}
\mbox{\epsfig{file=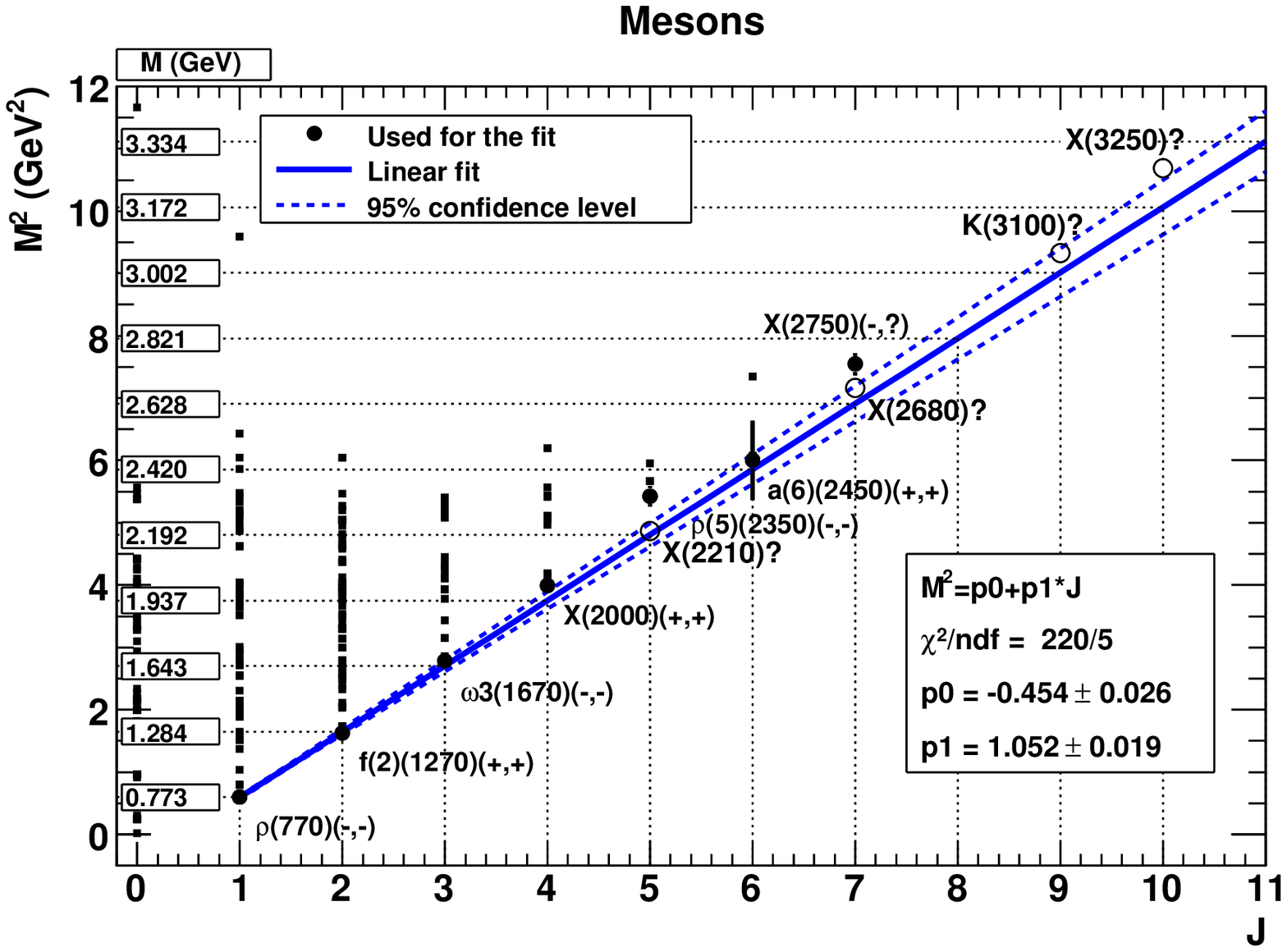,height=11cm}}
\caption
{
The ($J,M^2$) plane for mesons. The filled circles show the mesons with 
smallest masses and  $J\geq 1$
used for a  weighted-least squares linear regression (the solid line).
The dashed lines  indicate
a $95\%$ confidence-level interval for the linear regression.
For each meson used in the fit, 
space parity ($P$) and charge conjugation parity ($C$) are indicated in
parentheses.
Small filled squares show all other PDG
mesons and the open circles indicate guessed location
of several high-mass mesons with unknown $J$.
}
\label{fig3}
\end{center}
\end{figure}

\end{document}